\documentclass[twocolumn,showpacs,amsfonts,aps,prc,floatfix,superscriptaddress]{revtex4-1}

\usepackage[colorlinks=true, pdfstartview=FitV, linkcolor=red, citecolor=blue, urlcolor=blue]{hyperref}
\usepackage{amsmath}
\usepackage{bm}
\usepackage{graphicx}

\begin{document}

\title{Thermal photon $v_2$ with slow quark chemical equilibration}

\author{Akihiko Monnai}
\email[]{amonnai@riken.jp}
\affiliation{RIKEN-BNL Research Center, Brookhaven National Laboratory, Upton, NY 11973, USA}
\date{\today}

\begin{abstract}
Elliptic flow of direct photons in high-energy heavy ion collisions has been a topic of great interest as it is experimentally found larger than most hydrodynamic expectations. I discuss the implication of possible late formation of quark component in the hot QCD medium on the photon elliptic flow as quarks are the source of thermal photons in the deconfined phase. Hydrodynamic equations are numerically solved with the evolution equations for quark and gluon number densities. The numerical results suggest that thermal photon $v_2$ is visibly enhanced by the slow chemical equilibration of quarks and gluons, reducing the aforementioned problem.
\end{abstract}


\maketitle

\section{Introduction}
\label{sec1}
\vspace*{-2mm}

QCD matter under extreme temperatures has been studied extensively in high-energy nucleus-nucleus collisions. One of the most remarkable achievements in heavy ion physics is the realization of a deconfined phase called quark-gluon plasma (QGP) at the Relativistic Heavy Ion Collider (RHIC) in Brookhaven National Laboratory \cite{RHIC:summary} and the Large Hadron Collider (LHC) in European Organization for Nuclear Research \cite{Aamodt:2010pa}. The observed QGP is characterized with large azimuthal momentum anisotropy in hadronic particle spectra compared with the spacial anisotropy of the system originating from the collision geometry, which is believed to be the indication of the existence of a strongly-coupled medium near thermal equilibrium \cite{Kolb:2000fha}. Recent relativistic hydrodynamic analyses provide quantitative description of hadronic elliptic flow $v_2^h$ by including the effects of viscosity and geometrical fluctuation \cite{Schenke:2010rr}.
On the other hand, direct photon elliptic flow $v_2^\gamma$ turned out to be a few times larger than most of the hydrodynamic predictions in the experiments \cite{Adare:2011zr,Lohner:2012ct}. It should be noted that thermal emission of photons from anisotropic medium -- \textit{thermal photons} -- rather than photon production at the time of collision -- \textit{prompt photons} -- would be the origin of direct photon $v_2$ because experimental data indicate that the hot medium is chromodynamically opaque but electromagnetically transparent. There are no conclusive understanding of this discrepancy, though it has been approached from various prominent perspectives \cite{Chatterjee:2005de, Chatterjee:2008tp, Chatterjee:2011dw, Holopainen:2011pd, Chatterjee:2013naa, Chatterjee:2014nta, Liu:2009kta, vanHees:2011vb, Dion:2011pp, Shen:2013cca, Bzdak:2012fr,Muller:2013ila,Linnyk:2013hta}.

A heavy ion system at high energies goes through several stages after the collision. The color glass condensate picture \cite{McLerran:1993ni} indicates that the colliding nuclei is described as saturated gluons. It is then speculated to quickly isotropize, thermalize, and chemically equilibrate in a relatively short period of time $\sim \mathcal{O}(1)$ fm/$c$ after the collision. Since the elliptic flow, or momentum anisotropy, of the bulk medium is not fully-developed in the early stages of hydrodynamic evolution, $v_2^\gamma$ is na\"{i}vely expected to be smaller than $v_2^h$, which is calculated using the flow profile at the last stage of the evolution. 

In this paper, I investigate possible thermal photon $v_2$ enhancement by late quark chemical equilibration because thermal photons in the QGP phase are emitted by quarks. Most modern estimation for their emission rate is calculated for a completely equilibrated medium, though there is an indication that chemical equilibration of quarks and gluons could be slower than thermalization, a requirement for the onset of hydrodynamic flow, in several theoretical models \cite{Geiger:1992si,Biro:1993qt,collinearsplitting}. The photon emission from such systems is discussed extensively in terms of transverse momentum spectra \cite{Kampfer:1994rr,Strickland:1994rf}. It is naturally expected that $v_2^\gamma$ can become large if the thermal photons are created mainly in the middle-late stages where there already is a sizable momentum anisotropy in the medium. Thus late production of quark component in a heavy ion system can contribute to larger $v_2^\gamma$. 

The paper is organized as follows. Sec.~\ref{sec2} describes the formulation of relativistic hydrodynamics with the evolution of the quark and the gluon number densities. Thermal photon emission mechanism is embedded in the hydrodynamic model. In Sec.~\ref{sec3}, I present numerical results for the thermal photon elliptic flow for chemically equilibrated and non-equilibrated systems. Sec.~\ref{sec4} is devoted for discussion and conclusions. The natural unit $c = \hbar = k_B = 1$ and the Minkowski metric $g^{\mu \nu} = \mathrm{diag}(+,-,-,-)$ is used throughout this paper.

\section{Hydrodynamic model for quark-gluon mixture system}
\label{sec2}
\vspace*{-2mm}

In this section, I would like to develop a hydrodynamic model for chemically non-equilibrated systems with thermal photon emission. The time evolution equations for the quark and the gluon number densities are coupled to the energy-momentum conservation. The parton number densities are simply assumed to vanish at quark-hadron crossover. Thermal photon emission from the QGP and the hadron phases are considered for the estimation of photon elliptic flow. The purpose of the paper is to investigate qualitative implications of late chemical equilibration on the aforementioned quantity and precision analyses is beyond the scope of current discussion.

The background bulk medium for the thermal photon emission is the flow field provided by a hydrodynamic model. The energy-momentum conservation for multi-component hydrodynamics of quark-gluon mixture system in inviscid case would be given as
\begin{eqnarray}
\partial_\mu T_q^{\mu \nu} + \partial_\mu T_g^{\mu \nu} &=& 0, 
\end{eqnarray}
where
\begin{eqnarray}
T_q^{\mu \nu} &=& (e_q + P_q) u_q^\mu u_q^\nu - P_q g^{\mu \nu}, \\
T_g^{\mu \nu} &=& (e_g + P_g) u_g^\mu u_g^\nu - P_g g^{\mu \nu}. 
\end{eqnarray}
Here I make an ansatz that thermal equilibration is complete at the beginning of hydrodynamic evolution and assume the flows are common, \textit{i.e.}, $u^\mu \equiv u_g^\mu = u_q^\mu$. Then one can define the energy density $e = e_q + e_g$ and the hydrostatic pressure $P = P_q + P_g$. Here the subscripts $q$ and $g$ denote the quantities for quarks and gluons, respectively. The net baryon number is neglected assuming the numbers of quarks and antiquarks are equal in high-energy heavy ion systems. The matter-antimatter degrees of freedom are included in $T_q^{\mu \nu}$. 

The particle numbers are not conserved as the system is not chemically equilibrated. Since the parton number changing processes are (a) $g \rightleftharpoons g + g$, (b) $g \rightleftharpoons q+\bar{q}$, and (c) $q(\bar{q}) \rightleftharpoons q(\bar{q})+g$, the evolution equations for the quark and the gluon number currents for the 1-to-2/2-to-1 processes would be phenomenologically given as 
\begin{eqnarray}
\partial_\mu N_q^\mu &=& 2 r_b n_g - 2 r_b \frac{n_g^\mathrm{eq}}{(n_q^\mathrm{eq})^2} n_q^2 , \label{eq:dnqdt} \\
\partial_\mu N_g^\mu &=& (r_a - r_b) n_g - r_a \frac{1}{n_g^\mathrm{eq}} n_g^2 + r_b \frac{n_g^\mathrm{eq}}{(n_q^\mathrm{eq})^2} n_q^2 , \nonumber \\
&+& r_c n_q - r_c \frac{1}{n_g^\mathrm{eq}} n_q n_g \label{eq:dngdt},
\end{eqnarray} 
where $r_a$, $r_b$, and $r_c$ are the reaction rates for gluon splitting, quark pair production, and gluon emission from a quark, respectively. $n_q$ is the quark number density and $n_g$ is the gluon number density. The subscript ``eq" denotes the quantity in chemical equilibrium. The number currents can be decomposed as $N_q^\mu = n_q u^\mu$ and $N_g^\mu = n_g u^\mu$ in the inviscid case. The reaction rates are dependent on the temperature $T$ in thermal equilibrium. It should be noted that they are valid only above crossover temperature $T>T_c$ before quarks and gluons are confined in hadrons. Below the temperature the system is assumed to be hadronic, \textit{i.e.}, $n_q = n_g = 0$ and chemically equilibrated. One may in principle consider a chemical equilibrating hadronic system and write separate equations for each component but it is beyond the scope of this paper as our interest is the modification of thermal photon $v_2$ by the late quark production. 

The reaction rates are parametrized as $r_i = c_i T (i=a,b,c)$. Here $c_i$ is the dimensionless parameter that determines relaxation time scale. It is implied that $c_b > c_a, c_c$ leads to slower chemical equilibration when the initial medium is gluon rich, as would be the case for high-energy heavy ion systems.
Let us consider an extreme case where $c_a = c_c = 0$ for non-expanding systems, \textit{i.e.}, $u^\mu = (1,0,0,0)$. Then the analytic solutions for Eqs.~(\ref{eq:dnqdt}) and (\ref{eq:dngdt}) are given as
 \begin{eqnarray}
n_q (t) &=& 2 n_g (0) [1- \exp{(-r_b t)}], \label{eq:nq_approx} \\
n_g (t) &=& n_g (0) \exp{(-r_b t)} ,
 \end{eqnarray}
for $n_q \ll 1$ and $n_q (0) = 0$, which implies that $\tau_\mathrm{chem} \equiv 1/r_b$ could be defined as a typical time scale for chemical equilibration of the system. Note that the actual equilibration time can be longer due to the decrease of $r_b$ by the cooling effect of expanding systems and the presence of gluon number increasing processes. 
More detailed discussion which includes the quark recombination effect in semi-dense regions can be found in Appendix.

The parton gas picture yields $n_q^\mathrm{eq} = 3 \zeta (3) d_q T^3/4\pi^2$ and $n_g^\mathrm{eq} = \zeta (3) d_g T^3/\pi^2$ where $\zeta (3) \sim 1.20206$ and the degeneracies are $d_q = 24$ and $d_g = 16$ when $N_f = 2$. The color glass picture with the running coupling $\alpha_s = 0.2$ and the saturation momentum $Q_s = 2$ GeV suggests $n_g (\tau_0) \sim 5 Q_s^3 /6\pi^2$ where $Q_s$ is chosen so that $n_g (\tau_0) = n_g^\mathrm{eq} + n_q^\mathrm{eq}/2$ is satisfied. Here $\tau_0$ is the initial time.

Thermal photon emission rates from the QGP and the hadron phases are calculated based on Refs.~\cite{Arnold:2001ms} and \cite{Turbide:2003si}, respectively. Since the emission rates are functions of the quark distribution,  it is simply factored by $n_q/n_q^\mathrm{eq}$ for chemically non-equilibrated systems. The rate in the QCD crossover region is interpolated with a hyperbolic function as 
\begin{eqnarray}
E \frac{dR^\gamma}{d^3p} &=& \frac{1}{2}\bigg(1- \tanh \frac{T-T_c}{\Delta T} \bigg) E \frac{dR^\gamma_\mathrm{hadron} }{d^3p}\nonumber \\
 &+& \frac{1}{2}\bigg(1+ \tanh \frac{T-T_c}{\Delta T}\bigg) E \frac{dR^\gamma_\mathrm{QGP}}{d^3p} ,
 \end{eqnarray}
where the parametrization is set to $T_c=0.17$ GeV and $\Delta T = 0.017$ GeV. The photon emission below $p = 0.05$ GeV is cut off. The photon azimuthal particle spectra is given as 
\begin{equation}
\label{eq:dng}
\frac{dN^\gamma}{d\phi_p p_Tdp_T} =  \int dx^4 \frac{dR^\gamma}{d\phi_p p_Tdp_T},
\end{equation}
where $\phi_p$ is the azimuthal angle in momentum space and $p_T$ is the transverse momentum. The energy density is given by $p^\mu u_\mu$.
Then the differential photon $v_2$ is defined as the second harmonics in Fourier expansion:
\begin{equation}
\label{eq:v2gpt}
v_2^\gamma (p_T) = \frac{\int_0 ^{2\pi} d\phi_p \cos (2\phi_p ) \frac{dN^\gamma}{d\phi_p p_Tdp_T}}{\int_0 ^{2\pi} d\phi_p \frac{dN^\gamma}{d\phi_p p_Tdp_T}} .
\end{equation}

In this work, the hydrodynamic and the parton evolution equations are solved with a newly-developed (2+1)-dimensional hydrodynamic model, which assumes boost-invariance in the longitudinal direction. A smooth initial condition is considered for the present hydrodynamic analyses. The energy density profile is the one from Ref.~\cite{Kolb:2000sd} based on the wounded nucleon density with $b = 7$ fm for demonstrative purposes. The normalization is chosen so that $e_0 = 30$ GeV/fm$^3$ when $b = 0$ fm. The initial time is set to $\tau_0 = 0.6$ fm/$c$. The equation of state (EoS) in a chemically non-equilibrated system is non-trivial because most of the lattice QCD simulations are for equilibrated systems. 
Here it is simply approximated by interpolating the entropy density of the pion gas $s_\mathrm{hadron} = 3 \times 4 \pi^2 T^3/90$ and that of the parton gas $s_\mathrm{QGP} = 37 \times 4 \pi^2 T^3/90$ at $N_f = 2$ with the same hyperbolic function as the one for the photon emission rate.
This preserves consistency with the parton density picture introduced in the equilibrium values $n_q^\mathrm{eq}$ and $n_g^\mathrm{eq}$ for the deconfined phase.

\section{Results}
\label{sec3}
\vspace*{-2mm}

\begin{figure}[tb]
\includegraphics[width=3.2in]{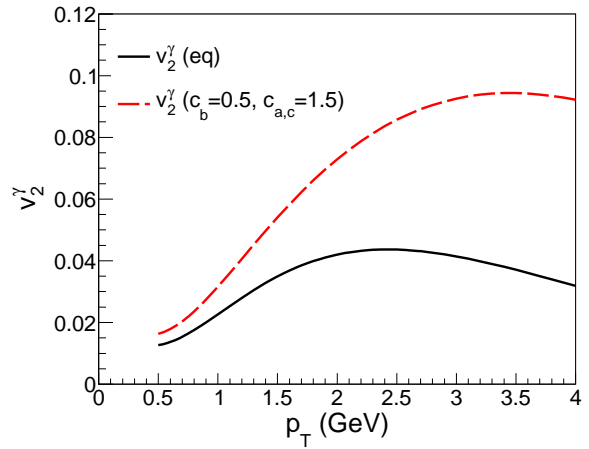}
\caption{The thermal photon elliptic flow $v_2^\gamma$ as a function of transverse momentum for a chemically-equilibrated medium (solid line) and a non-equilibrated medium with $c_a = c_c = 1.5$ and $c_b = 0.5$ (dashed line). }
\label{fig:1}
\end{figure}

The differential photon elliptic flow $v_2^\gamma (p_T)$ in chemically non-equilibrated QCD medium is shown in comparison with the one in chemical equilibrium for the transverse momentum range $0.5 < p_T < 4$ GeV in Fig.~\ref{fig:1}. The reaction rate parameters are chosen as $c_a = c_c = 1.5$ and $c_b = 0.5$ to represent late chemical equilibration of quarks. The quark number density at the initial time is set to vanishing. The contribution of thermal photon emission until the hadronic freeze-out at $T_f = 0.15$ GeV is taken into account. 

One can see that the elliptic flow is enhanced for the late production of the thermal photon emission sources. Considering that most of the hydrodynamic analyses so far under-predicted photon momentum anisotropy, the result would provide an important viewpoint on how to interpret the phenomenon in terms of pre- and post-equilibrium physics, \textit{i.e.}, quark pair production in splitting-based picture, \textit{e.g.}, Ref.~\cite{collinearsplitting} and relativistic hydrodynamics. The mean transverse momentum of thermal photons is also enhanced. 
One has to be careful because the results are sensitive to the EoS, the duration time of hydrodynamic phase, and the initial conditions. It should be noted that validity of background hydrodynamic flow analyses should not be na\"{i}vely assumed for mid-high $p_T$ regions above $\sim 2$ GeV. Also effects of viscosity in hydrodynamic flow and distribution functions for quark-gluon/hadronic particles would both become important in those mid-high $p_T$ regions.

\begin{figure}[tb]
\includegraphics[width=3.2in]{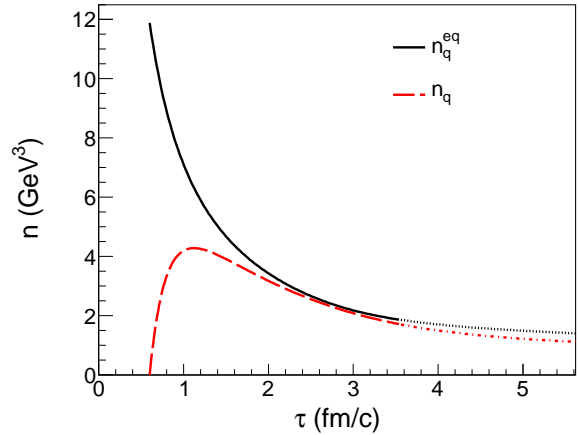}
\caption{The quark number densities in chemical equilibrium (solid line) and in dynamical evolution with $c_a = c_c = 1.5$ and $c_b = 0.5$ (dashed line) until the local temperature reaches $T_c = 0.17$ GeV. The dotted and the dash-dotted lines denote that the system is near the crossover region $T < T_c + \Delta T$ where $\Delta T = 0.017$ GeV.}
\label{fig:2}
\end{figure}

Magnitude of chemical equilibration of the system at the center of the hot medium $x = y = 0$ fm is shown in Fig.~\ref{fig:2} before the system reaches $T = 0.17$ GeV where the parton picture would no longer be valid. The longitudinal expansion of the system leads to quick reduction in the number density during the time evolution. The system eventually approaches the one in equilibrium. The effective relaxation time for chemical equilibration is $\sim 2$-$3$ fm/$c$ with the current choice of reaction rate parameters, which is roughly in agreement with the estimation with Eq.~(\ref{eq:nq_approx}) because $\tau_\mathrm{chem} = 1/c_b T \sim 2$ fm/$c$ for $c_b = 0.5$ and $T \sim 0.2$ GeV. One sees deviation from $n_q^\mathrm{eq}$ below the crossover region because the EoS is no longer from the one for the parton gas model, representing the breakdown of the quark-gluon picture.

\begin{figure}[tb]
\includegraphics[width=3.2in]{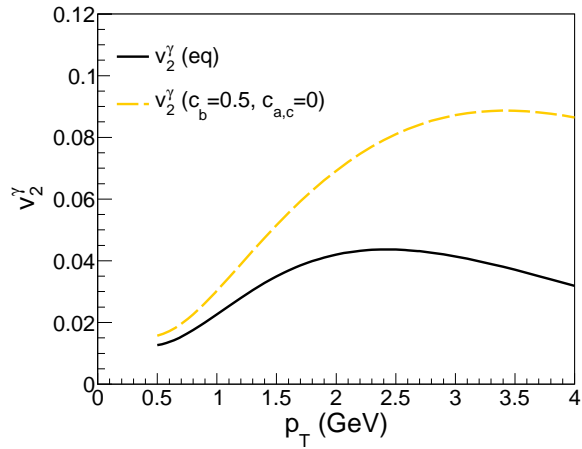}
\caption{The thermal photon elliptic flow $v_2^\gamma$ as a function of transverse momentum for a chemically-equilibrated medium (solid line) and a non-equilibrated medium with $c_a = c_c = 0$ and $c_b = 0.5$ (dashed line). }
\label{fig:3}
\end{figure}

Fig.~\ref{fig:3} shows the case where only the quark production process is present, \textit{i.e.}, $c_a = c_c = 0$ and $c_b = 0.5$ for demonstration of the effects of the parton number evolution mechanism on thermal photon $v_2^\gamma$. The result again shows visible enhancement and is also not so different from the previous calculation with gluon emission processes. This can be interpreted that the chemical relaxation time scale is mostly determined by the slowest process. By choosing larger quark production rate $r_b$, \textit{i.e.}, setting the shorter chemical relaxation time, one recovers the result in equilibrium.

\section{Discussion and Conclusions}
\label{sec4}
\vspace*{-2mm}

I have developed a hydrodynamic model with quark and gluon number density evolution and estimated thermal photon emission from the expanding medium with boost invariance at mid-rapidity. The numerical results show that elliptic flow for thermal photons can be visibly enhanced by the slower quark chemical equilibration. This can be one of the mechanisms for explaining the large photon $v_2$ problem in high-energy heavy ion collisions \cite{Adare:2011zr,Lohner:2012ct}.
There is possible overestimation in the current analyses because the quark number should be finite at the beginning of the hydrodynamic stage while it is simply assumed to be vanishing in the present estimation. Also more microscopic treatments would be necessary for determining the parameters for the chemical equilibration processes without ambiguities.

It should be noted that late quark production leads to the reduction in photon $p_T$ spectra. Since the spectra is often employed to experimentally constrain the temperature of the medium \cite{Adare:2008ab,Wilde:2012wc}, this would lead to higher medium temperature for chemically-equilibrating systems for consistency. Introduction of more realistic off-equilibrium equation of state and fine-tuning of initial conditions, including thermalization times, would be essential for more quantitative discussion.

Further future prospects include estimation of the effects of shear and bulk viscosities, especially since it is known that in addition to the flow modification during the hydrodynamic evolution, the distortion of distribution can have non-trivial effects on particle spectra and elliptic flow of hadrons at freezeout \cite{Teaney:2003kp, Monnai:2009ad,Denicol:2009am}. 
Triangular and higher-order flow of thermal photons $v_n^\gamma$ are interesting observables because they would play an important role in distinguishing the origin of the large elliptic flow for photons, \textit{i.e.}, one should be able to distinguish if it is based on internal medium properties of QCD systems as discussed in the present paper or on external properties of heavy ion geometry such as strong magnetic fields. One can expect that the former scenario would lead to large $v_3^\gamma$ while the latter would not. Also if one considers separate flows for quarks and gluons, then interesting interplay of the intrinsic medium properties and the heavy ion geometry should be observed as only the former would be directly affected by the strong magnetic field.

\begin{acknowledgments}
The work is inspired by fruitful discussion with B.~M\"{u}ller. The author would like to thank for valuable comments by Y.~Akiba and L.~McLerran on the paper.
The work of A.M. is supported by RIKEN Special Postdoctoral Researcher program.
\end{acknowledgments}

\appendix*
\section{ANALYTIC EXPRESSION OF QUARK AND GLUON NUMBER DENSITIES}
\label{appA}
\vspace*{-2mm}

The equations (\ref{eq:dnqdt}) and (\ref{eq:dngdt}) can be analytically solved without neglecting the quark recombination terms for non-expanding geometry when $c_a = c_b = 0$. The quark number and gluon number densities at a given time $t$ are
\begin{eqnarray}
\label{eq:nq_rec}
n_q &=& \frac{(n_q^\mathrm{eq})^2}{4 n_g^\mathrm{eq}} \bigg[I_1 \tanh{\bigg( \frac{I_1}{2} r_b t + I_2 \bigg) } -1 \bigg], \\
n_g &=& n_g^\mathrm{eq} + \frac{n_q^\mathrm{eq}}{2} - \frac{(n_q^\mathrm{eq})^2}{8 n_g^\mathrm{eq}} \bigg[I_1 \tanh{\bigg( \frac{I_1}{2} r_b t + I_2 \bigg) } -1 \bigg], \nonumber \\
\end{eqnarray}
where
\begin{eqnarray}
I_1 &=& 1 + \frac{4n_g^\mathrm{eq}}{n_q^\mathrm{eq}} , \\
I_2 &=& \tanh^{-1} \frac{1}{I_1},
\end{eqnarray}
when $n_q (t=0) = 0$.
One has $I_1 \sim 4.55556
$ for parton gas with $N_f = 2$, which implies that the typical time scale for equilibration in the presence of the recombination process would be slightly shorter the inverse reaction rate $1/r_b$ in non-expanding medium. This would be because equilibration does not require complete conversion of gluons into quarks as is the case for systems with no recombination. The actual chemical equilibration would become longer than this estimation because of the gluon number increasing processes for non-vanishing $c_a$ and $c_c$.

\bibliography{basename of .bib file}

\end{document}